# QUANTUM CRITICISM: A TAGGED NEWS CORPUS ANALYSED FOR SENTIMENT AND NAMED ENTITIES

Ashwini Badgujar[1], Sheng Cheng[1], Andrew Wang[1], Kai Yu[1], Paul Intrevado[2] and David Guy Brizan[1]

[1]Department of Computer Science
University of San Francisco, San Francisco, CA, USA
{abadgujar, schen127, awang27, kyu18}@dons.usfca.edu, dgbrizan@usfca.edu
[1]Department of Mathematics and Data Science
University of San Francisco,San Francisco, CA, USA
pintrevado@usfca.edu

## ABSTRACT

*In this project, we continuously collect data from the RSS feeds of traditional news sources. We apply several pre-trained implementations of named entity recognition (NER) tools, quantifying the success of each implementation. We also perform sentiment analysis of each news article at the document, paragraph and sentence level, with the goal of creating a corpus of tagged news articles that is made available to the public through a web interface. Finally, we show how the data in this corpus could be used to identify bias in news reporting.*

## KEYWORDS

*Content Analysis, Named Entity Recognition, Sentiment Analysis*

## 1. INTRODUCTION

Many of us implicitly believe that the news we consume is an important summary of the events germane to our lives. Regardless of how we divide ourselves—by demographics, political leaning, profession or other socioeconomic schism—we rely on trusted individual journalists and the news organizations to which they belong to distill stories and provide unbiased context.

There are several organizations that attempt to address this need. USAFacts.org is a non-profit organization and website which offers a non-partisan portrait of the US population, its government's finances, and government's impact on society. Similar sites and outlets have had the same mission, perhaps most prominently MIT's Data USA and the US government's data.gov. These efforts, however, largely deal with quarterly or bi-annual government reports, excluding day-to-day news analysis about business, politics, etc.

More timely news on these excluded topics can typically be found reported on by private news organizations, often funded by a subscription or ad-based model. There are, however, a subset of articles that are freely available to the public. News producers often promote selected articles through their real simple syndication (RSS) feeds, consumed by phone or web applications such as Feedly, NewsBlur and FlowReader, among others.

News organizations should be a reflection of the populations they represent. Yet, despite ease of access to news articles through RSS feeds, we find a dearth of resources supporting the analysis of said news articles, e.g., how the news is reported or how it may be affecting our lives over time. For example, observing climate change denial, one journalist from Vox, David Roberts, has named the current American philosophical divide "tribal epistemology," specifically discussing the tribalism of information through the news [1]. While his presentation is compelling, the idea of tribal epistemology is largely delivered without an analysis of the news from sources which he critiques. Roberts's lack of analysis could be the result of having no facile manner to find and analyse daily news articles from multiple sources in a single corpus.

In our survey of existing news corpora (Section 2), we find existing corpora lacking in one or more aspects, including cost, availability, coverage and/or analysis. We therefore create our own corpus, Quantum Criticism, to address these issues. Specifics of the tools and approaches we use to build our corpus are discussed in Section 3. We discuss the performance of our tools in Section 4. We aspire for our corpus to be used by journalists and for those in academic research to establish trends, identify differences, and affect change in news reporting and its interpretation. In Section 5, we demonstrate two ways in which our corpus can be used to uncover potential media bias.

## 2. RELATED WORK

We begin the Related Work section by highlighting existing corpora that have some coverage or analysis limitation, discussed in Section 2.1. Section 2.2 briefly reviews common tasks in natural language processing, as well as some of the available tools for accomplishing those tasks. Lastly, in Section 2.3, we explore several use-cases of existing news corpora.

### 2.1. Corpora

There are several outcomes of forming a news-based corpus. One may be the task of language modelling. Journalists and news organizations can be barometers of when a word gets introduced to a language. Another important use of news-based corpora is the derivation of larger social patterns from individual units of reporting.

The consumers of a news corpus must regard journalists and news organisations as imperfect messengers. As far back as 1950, White [2] demonstrated that the news we read is frequently collated by a set of "gate keepers" who filter candidate events. These gate keepers may have biases based on ideological (liberal or conservative) leanings, race or gender [3], economic interdependence [4] and geopolitical affiliation [5], likely only some of the many factors influencing a news story's selection. One use of a properly constructed corpus could be the unearthing of selection bias or other biases.

Selection bias may be the result of the choices of not only the specific journalists but also the news organizations and their owners [6]. In a large-scale study based on articles from the GDELT database, [7] lays out the constraints under which the news organizations operate and quantify the selection bias of news organizations.

Prior to building our Quantum Criticism corpus, we considered a number of other corpora assembled from news articles, all appearing online. The Linguistic Data Consortium (LDC) has an extensive collection, including the New York Times corpus [8], which we use for validation of our tools. (Details in Section 4.) This corpus contains 1.8 million news articles from the New York Times over a period of more than 10 years, covering stories as diverse as political news and restaurant reviews. Articles are provided in an XML format, with the majority of the articles tagged for named entities—persons, places, organizations, titles and topics—so that these named entities are consistent across articles.

The LDC also offer the North American News Corpus [9], assembled from varied sources, including the New York Times, the Los Angeles Times, the Wall Street Journal and others. The primary goals of this corpus are support for information retrieval and language modelling, so the count of "words"—almost 350 million tokens—is more important than the number of articles.

Also offered by the LDC is the Treebank corpus [10], often called the Penn Treebank, which has been an important and enduring language modelling resource; see [11] for an early use of this corpus, and [12] for a more recent implementation.

Collectively, the LDC corpora and their like are excellent resources for news generated from a discrete number of sources during a particular period of time. Because of their volume of articles and tokens, and because they are mostly written in Standard American English, they are ideal for building language models from the period during which they were collected.

However, we find the aforementioned corpora broadly lacking in a number of areas, chiefly, in their static nature: these corpora do not continuously collect new articles. Depending on the research being conducted, researchers may require current articles as well as historic ones. We also find flaws in the tagging of the articles in the New York Times Annotated Corpus, but leave the full treatment of this to Section 4. Finally, we find that processing these articles requires a non-trivial cost and effort. Finding articles in which a particular person, place or organisation is mentioned requires a search through a considerable number of articles, for which there are no additional tags.

In contrast to the offerings by the LDC, the Global Database of Events, Language and Tone [13], known as GDELT, has a dizzying array of tools for searching and analysing their corpus. With a public, no-cost access to articles from 1979 to present, albeit offered at a 48-hour delay, and a commitment to the continued collection of news from a wide variety of sources, GDELT's offerings have resulted in insightful results, some of which are explored herein.

One criticism of GDELT by Ward et al. [14] is that the collection effort has been optimized for volume of news articles and speed of analysis through automated techniques, sacrificing the careful curation of articles. This results in the improper classification of articles, erring mostly toward false positives, i.e., presenting more news articles as related to an event than is warranted.

In terms of implementation, our Quantum Criticism cor-pus is closest to the News on the Web (NOW) Corpus, itself a public-facing version of the Corpus of Contemporary American English [15]. As of the time of this writing, this corpus reports containing 8.7 billion words from a number of American English sources, including such varied sources as the Wall Street Journal and [tigerdroppings.com](tigerdroppings.com), the student newspaper of Louisiana State University.

While the diversity of our Quantum Criticism corpus is not as extensive as what we find in the NOW Corpus, our initial version of the Quantum Criticism corpus contains one non-American English source and allows the user to specify the source(s) for a query. We believe the power of our search and presentation makes our corpus a better analysis tool.

### 2.3. Overview of NLP Tools

We analyse news articles in two ways: through named-entity recognition and sentiment analysis. Our search tool exposes the results of these analyses simultaneously.

In free-form text, named-entity recognition (NER) seeks to locate and classify the names of (among other entities) people, organisations and locations. Although there are other possible categories of named entities, we selected these three classes based on available resources and commonality of model outputs. Three powerful and oft used NER tools include BERT (Bidirectional Encoder Representations from Transformers, [16]), which uses BIO tagging, CoreNLP [17], which offers both IO and BIO tagging, and spaCy [18], which employs IOB tagging.

### 2.3. Use Cases of Corpora

Using corpora and NLP tools, we can discover the biases of a journalist, a news organisation or the target audience of the news. The effects of biases can effect change on the political or sociological lives of a people. We see some interesting examples of these effects. While work by Rafail and McCarthy [19] stops short of the claim that some news organizations made the Tea Party—a small, right-leaning movement—a political force, there may be ample evidence to draw such a conclusion. The suggestion is that the news media simplified the message of the party so that it could be consumed by a wider audience, as well as amplified the coverage of the party's events beyond the size its supporters would normally warrant given their numbers.

A more pernicious effect may be seen in the coverage of the Persian Gulf "Crisis" and subsequent war of the early 1990s [20]. Here, the media was focused on stories which, among other effects, made readers inclined to favour military rather than diplomatic paths. In turn, this had an effect on the political leadership of the time. The authors also find an interesting effect wherein the selection bias for stories was proportional to public interest in such stories. Interestingly, work by Soroka et al. [21] suggests the opposite effect may be a force. Here, the "strength" of sentiment in social media reactions differ from the news media coverage in some economic news coverage. As a result, the contexts and degree to which public opinion affects news coverage or vice versa deserve additional study.

Systematic analysis of media coverage often involves framing the content from the point-of-view of the reader. A paper by An and Gower [22] discusses five frames (attribution of responsibility, human interest, etc.) and two "responsible parties" (individuals vs. organizations) in coverage of crises, finding that some frames are more common than others. Similarly, Trumbo [23] examines the differing reactions of scientists and politicians to climate change. While analysis approaches tend to focus on the content produced, work by Ribeiro et al. [24]

examines the political leanings and demographics of the target audience through the advertising associated with the content. We see this kind of side-channel investigation as promising, especially if applied systematically to a large set of data.

Our Quantum Criticism corpus is designed with these types of analysis in mind. We tag each article for named entities and sentiment and expose this corpus to the public. We expect this corpus to have multiple purposes, including sociological research on influential people and organizations, "framing" news articles and assigning responsible parties, and the detection of selection bias and other biases in a media organization's coverage. We provide details on how each element of our pipeline is built, and quantify the performance using well-established metrics. We conclude by validating the tools employed and discussing two use cases for our corpus.

## 3. CORPUS AND DATA PROCESSING

The data used for our Quantum Criticism effort was collected, managed, and processed using a proprietary system designed to scrape, parse, store and analyse the content of news articles from a variety of sources. Several sentiment and named entity recognition tools were run against the collected news articles. We also implemented a custom entity resolution algorithm, providing a rich data set upon which to explore several hypotheses. A pictorial summary of the ingestion, analysis and storage pipeline is shown in Figure 1.

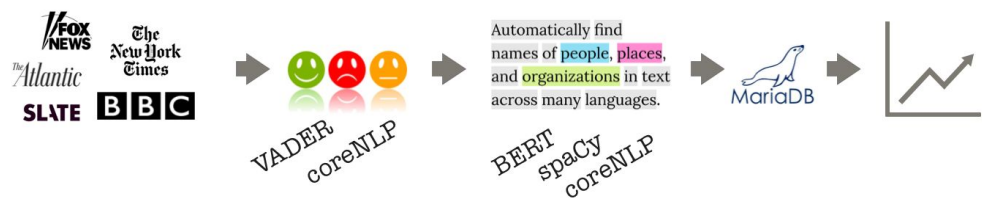

**Figure 1: A Summary of the Ingestion, Analysis and Storage Pipeline**

### 3.1. News Scraper

Several custom web scrapers were created for retrieving news articles from various online news organizations. All web scrapers were run every two hours to retrieve articles from the following five news sites: the Atlantic, the British Broadcasting Corporation (BBC) News, Fox News, the New York Times and Slate Magazine. Web scrapers continue to run every two hours in perpetuity, scraping additional news articles. Collectively, the web scrapers used each news organization's RSS feed as input, storing the scraped output into a custom database. Article URLs were used for disambiguation; where two scraped articles shared a URL, the most recently retrieved article replaced previous versions of articles.

As of November 2019, we collected a total of 105,000 news articles from five media organizations. Figure 2 depicts the number of cumulative articles scraped for each news organization over time. Even though articles from Fox News were regularly scraped four months later than other news sources, the number of articles scraped rose quickly, and now constitutes the news organization with the most scraped articles. Given the news scrapers run at regularly scheduled two-hour intervals for all news organization, this suggests that Fox News updates its RSS feed with new articles far more often than others, and the Atlantic updates its RSS feed far less frequently than others.

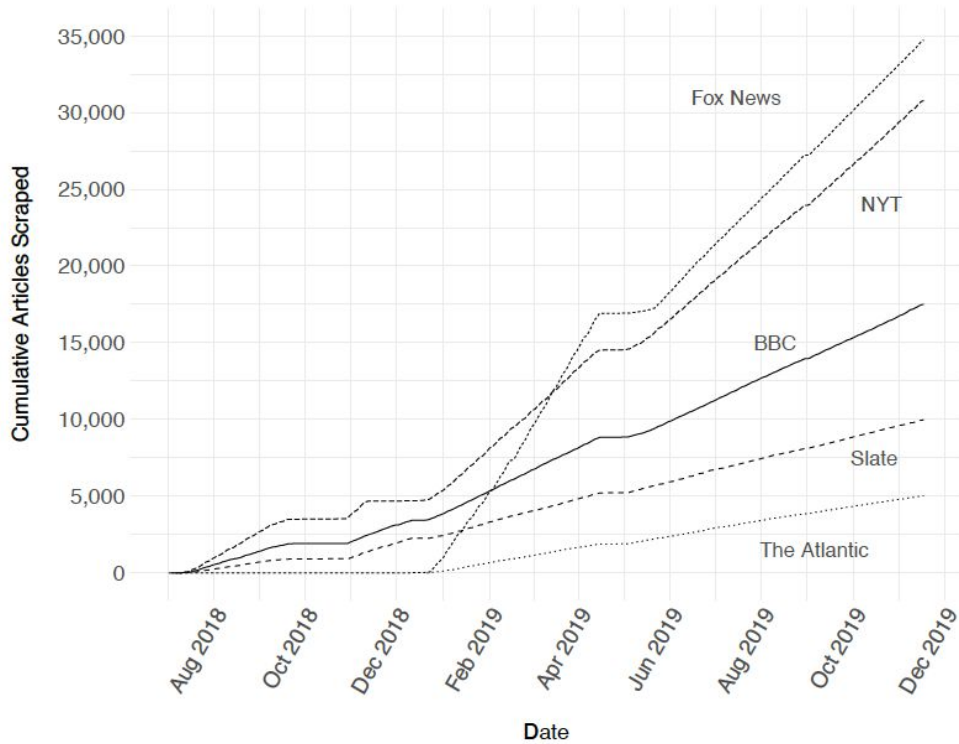

**Figure 2: Cumulative Quantity of Articles Scraped by News Organization**

### 3.2. Data and Database Management

All scraped data is stored in a MariaDB relational database. We considered a NoSQL database, especially one focused on storing documents, such as MongoDB; however, we found that a relational database was appropriate for the needs of this project.

We constructed many "primary" tables to support the scraped articles. The most important of these tables are the article, media (e.g., The Atlantic, BBC, etc. representing the news organization) and entity (a named person, location or organization) tables. To support modelling the many-to-many relationship between article and entity, we have one "join" table (article entity). To support the work in sentiment analysis and named entity recognition; we also created tables to store the outputs of the algorithms for these tasks. For sentiment analysis, we created a table called "sentiment. For named entity recognition, we created a table "entity." Other tables in our schema are omitted for brevity. Courtesy of dbdiagram.io, a schema appears as Figure 3.

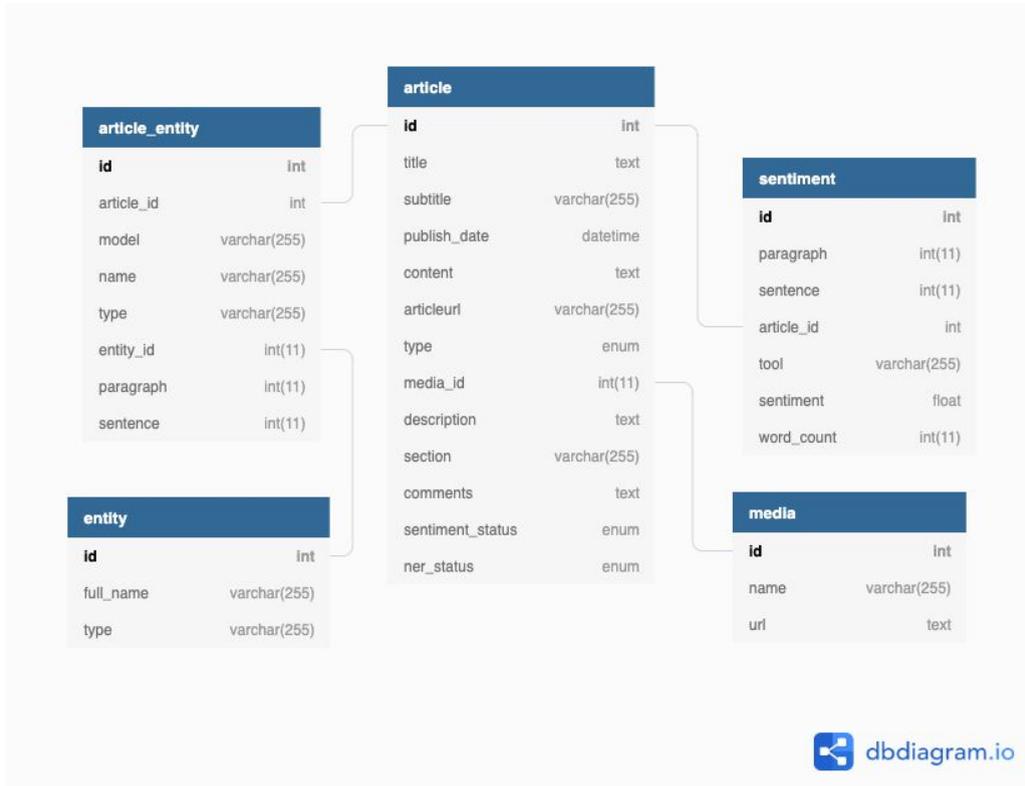

Figure 3: Schema for the Quantum Criticism Database

### 3.3. Sentiment Analysis

For each news article, we generated a sentiment score. We employed both the VADER (Valence Aware Dictionary and sentiment Reasoner) [25] module, as implemented in NLTK [26] in python, as well as CoreNLP sentiment analysis. Sentiment scores in VADER are continuous values between -1 (very negative) to +1 (very positive), with 0 representing neutral sentiment. Sentiment scores in CoreNLP are integer values between 0 (very negative) and 4 (very positive), with 2 representing neutral sentiment.

Sentiment analysis tools were run against each sentence and each paragraph in the article, as well as on the entire article. For example, if an article contained two paragraphs, where paragraph 1 contains two sentences and paragraph 2 contains one sentence, we would have calculated six different sentiment scores per sentiment analysis tool: one for each sentence (3), one for each paragraph (2), and one for the article (1). This deconstructed approach allows researchers to associate named entities with their associated sentiment at a quantum level. This granular level of sentiment may help disambiguate the sentiment of an article with respect to the named entities. For example, an article from a conservative news organization may be positive overall, however, it is likely to be more critical of more liberal politicians, organizations or causes mentioned therein, and more supportive of conservative organizations or causes. Our quantum approach to sentiment analysis allows researchers to parse sentiment at the sentence level and associate that sentiment with named entities, independently of the paragraph or article in the aggregate.

### 3.4. Named Entity Resolution

We employed eight named entity recognition (NER) models from CoreNLP, spaCy, and BERT packages to identify PERSONs, ORGANIZATIONs and LOCATIONs in news articles. While some models predict different NER categories, we sought only those entities which were tagged as above.

We store each NER model output individually in our database. In many articles, a named entity is referenced by a complete name or title, and subsequently, by a shortened version. For example, a recent New York Times Opinion article ([Democrats' Vulnerabilities? Elitism and Negativity](#)) first refers to politician Alexandria Ocasio-Cortez by her full name, and then subsequently as Ocasio-Cortez. In order to connect references to the same named entity, we implemented a custom entity resolution algorithm. Owing to the highly structured manner in which we observed news articles were written, we expected to observe the pattern of an entity's full name, followed by partial name. Our algorithm therefore matched any name extracted in an article as a substring, to the most recent instance of another name in the same article. Where a match occurred, the two names are determined to be the same entity. Such an entry is matched or created by category (PERSON, LOCATION, ORGANIZATION).

This process often failed for abbreviations such as the acronym F.B.I.—in reference to the Federal Bureau of Investigation—with periods left out, resulting in FBI. We therefore also created custom code to query a corpus of abbreviations and associate acronyms to their full names. Only full names were stored in the database. We label each such instance of the full name a resolved entity. The entity resolution algorithm is depicted in Figure 4. Entities are also resolved across articles in a similar manner.

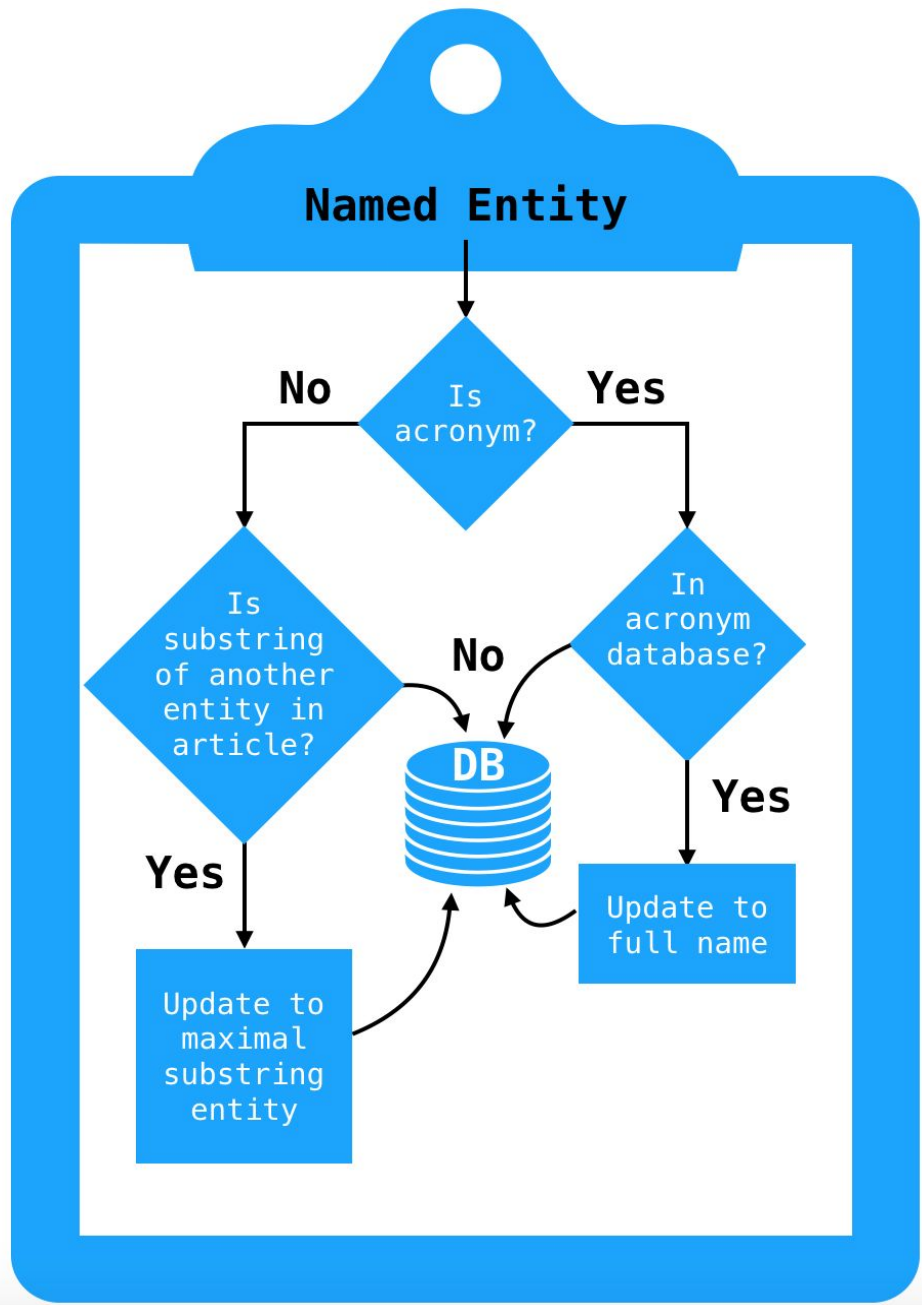

**Figure 4: Pseudo-code for Named Entity Resolution Within a News Article**

### 3.5. Web Interface

We designed and implemented a web interface for our corpus. Through this interface, a user can specify basic search criteria for the articles, specifically: the entity name, in whole or part, of the entity to be searched; the news source(s) to be searched from among those in our database; and the date or date range of the articles. (See Figure 5.)

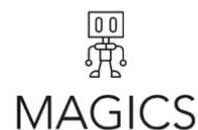

Figure 5: Search Screen for Web Interface

Additionally, advanced search criteria—not shown in Figure 5 but available on the live web interface—allow the user to include additional filters for specific NER tools, the sentiment tools, and/or the level of granularity (article, paragraph or sentence) to be reported.

Upon executing a successful query, a small subset of the results is displayed so that the user may perform a quick validation. In addition, a link is provided to allow the user to download the full set of results in comma-separated value (CSV) format. Each row in the result set contains the fields listed in Table 1.

Table 1: Fields in the Result Set of the Web Interface

| Field | Type | Notes |
| --- | --- | --- |
| id | integer | Database table ID |
| entity | string | Full name of entity |
| entity id | int | Database table ID |
| type | Enum | PER, LOC or ORG |
| date | Date | Last modified date |
| url | String | Article's URL |
| NER tool | String | Name of the NER tool |
| paragraph | int | Possibly NULL or empty |
| sentence | int | Possibly NULL or empty |
| sentiment score | float | |
| sentiment tool | String | Name of the model |
| media name | String | "Fox News", "Slate," ... |
| media url | String | URL of the news org. |

Notably absent from the columns in the result set are the contents of the article. This absence is deliberate. While recent legal rulings have suggested that distributing content produced by third parties is permissible, we are unsure about whether that ruling is the final word on this or whether the ruling applies globally. As a result, we provide the URL to the source article, allowing the user to download the content themselves.

# 3. VALIDATION

We employ well-studied tools with established performance benchmarks in our data ingestion and processing pipeline. In this section, we describe how we evaluated the performance of those tools. For the validations reported in this section, we used two news corpora: a historical New York Times corpus and our Quantum Criticism corpus of scraped news articles.

## 4.1. Named Entity Recognition Validation

We test the efficacy of our eight NER models across three different NER tools using two approaches. Firstly, we executed each of the models against the articles in the New York Times Annotated Corpus with a 1st of December publication date across all 20 years covered by the corpus. Secondly, we explore the fidelity of the NER tools by examining how well they identify the 538 members of the U.S. Congress (Senate and House of Representatives), as identified by [Ballotpedia](#).

### 4.1.1. NER Tools vs. the LDC New York Times Corpus

To determine the fidelity of our results, we ran each NER model against the New York Times Annotated Corpus [8], for which named entities are provided as an adjunct list. The corpus contains 1.8 million articles from the New York Times from the years between 1987 and 2007.

While we found that the 4,713 articles from the 1st of December, 1987–2007 was a sufficiently ample volume from which to draw conclusions, we tested an additional ten months of data for the spaCy and CoreNLP models, finding no significant deviation from the results we report here.

For each of the articles published on the 1st of December, we determined the mean (and standard deviation) of the fol-lowing in each article: the number of tokens per article: 587.2 (643.7); and the number of named entities identified by the models per article: 31.8 (43.9). For each of the models, we also computed the precision, recall and F1 score for each article. The BERT `bert.base.multilingual.cased` model generated the highest mean precision and mean F1 scores of 0.1753 and 0.2549, respectively, whereas the highest mean recall score was obtained from the CoreNLP `en-glish.all.3class.distsim.crf.ser` model.

We observed a consistently low F1 score for all NER mod-els, despite the variable number of entities identified by the classifiers. Some of this poor performance may be ex-plained by the models' generation of improperly resolved entities in the body of the article. However, we believe that this poor performance can be largely attributed to errors in the labels of the source corpus.

To confirm this hypothesis, we examined several articles from the New York Time Annotated Corpus, and found dis-agreement with the named entities identified in the manual tagging of the corpus. Filtering for the named entity classes PERSON, LOCATION and ORGANIZATION in one of these examined articles, *[Homicides Up in New York; Other Crimes Keep Falling](#)*, we find only three tags from the corpus: Cara Buckley, the article's author; New York City, the location being reported; and the Federal Bureau of Investigation. These instances identified by the corpus are highlighted in blue. In contrast, one of the authors, a native English speaker who has performed several annotation tasks on other projects, identified several other named entities. These additional named entities are highlighted in yellow. Interestingly, two of the models we use, `bert.base.cased` and `bert.base.multilingual.cased`, added a spurious named entity label in this case, "Homicides." We highlight this deviation in red.

In the articles we inspected, our annotator found that his identification is closer to the results we obtain from the NER models. Interestingly, BERT appears to exhibit a tendency toward combining tokens to form named entities (e.g., "Ms. Pickett" with "Fort Greene" to form "fort green pickett"), and toward listing names of people with family name first (e.g., "pickett, cheryl"); we converted to the surname-last ordering common in American English. While we believe the output of the BERT models is closer aligned to our expectations, it also clearly misidentified tokens (e.g., "homicides" in the above example) as named entities and misclassified named entities, commonly determining a person's name to be a LOCATION instance.

Figure 6: Labels identified by the NY Times Corpus (blue) and additional labels identified by our annotator (yellow) and spurious labels identified by BERT (red)

Although our manual evaluation of the corpus was limited in scope, it does lead us to believe that named entities are generally under-reported by the corpus. We therefore concluded that the most reliable metric in comparing our mod-els to the corpus is recall. This allows us to treat each NER system as a detector, i.e., to determine what fraction of the entities in the articles' annotations are identified by the NER models.

### 4.1.2. NER Tools vs. the Quantum Criticism Corpus

Because we were unable to label an extensive set of articles, we have no ground truth for the performance of the NER tools against our corpus. Instead, we largely rely on the metrics of the NER tools against the NY Times Corpus for this. However, we are interested in determining the number or rate of misclassifications of entities.

For this, we used the 538 current members of the U.S. legislative branch (Congress) as identified by Ballotpedia. This list includes all members of the U.S. Senate and the U.S. House of Representatives. Of these, 372 (69%) are mentioned at least once in the articles we scraped. We examined these mentions as a way to assess the quality of the NER tools employed.

Looking across all eight NER tools from BERT, CoreNLP and spaCy, 96.9% of all entities resolved to the correct classification of PERSON. There were, however, some no-table deviations. Two models however, one from spaCy and another from CoreNLP, consistently misidentified Congress people as ORGANIZATION instances, at a rate of 5.06% and 2.98%, respectively, as depicted in Figure 7. This behaviour may be, in part, be attributed to the fact that Congresspeople often lead or participate in important organizations, and are therefore often conflated with them. For example, Nancy Pelosi, the former and current Speaker of the U.S. House of Representatives at the time of this writing, is often misidentified as an ORGANIZATION given her leadership role. Perhaps more interesting is the performance of the `english.conll.4class.distsim.crf.ser` model in CoreNLP, which misidentifies 7.76% of all Congresspeople.

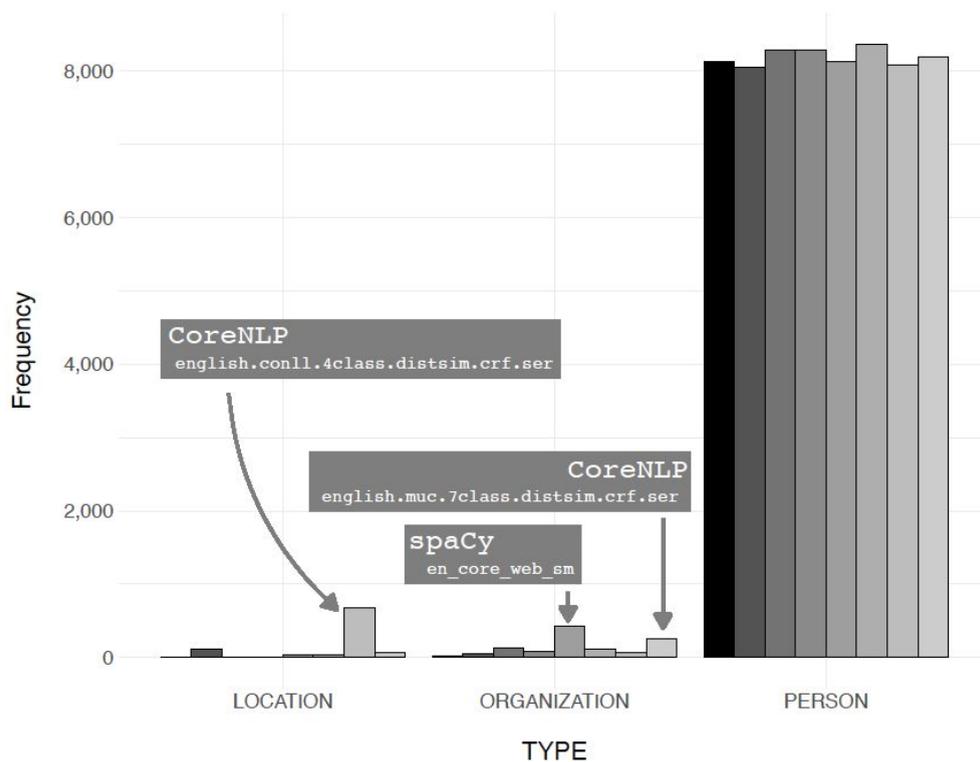

Figure 7: (Mis)Classification of PERSON Entities by NER Model

**4.2. Entity Resolution Validation**

Querying our scraped articles, we sought to explore how well our proprietary entity resolution algorithm worked to resolve the names of Congresspeople. We searched our scraped data using each space-separated or hyphen-separated token from the full names of members of Congress listed on Ballotpedia. These results were then manually checked to retain only valid references to the individuals in question.

To illustrate the above, we use Nancy Pelosi, who is the most-mentioned Congressperson in our database. Using the strings "%Nancy%" and "%Pelosi%" as our search criteria and removing unassociated entities (e.g., "Pino Pelosi", "Nancy Reagan", etc.), we identified thousands of references to 475 entities. "Nancy Pelosi" as a PERSON instance is the most common entity, with 1,915 scraped articles. "Pelosi," misidentified as an ORGANIZATION 371 times, is the second most frequent occurrence. "Ms. Pelosi," "Pelosi" and other variants are less frequent. Figure 8 shows the top ten entities for Nancy Pelosi along with the frequency with which they occur. We can measure the precision of our model with respect to an individual instance as the most frequent occurrence. In this case, Nancy Pelosi (PERSON) represents 53% of all the references to her.

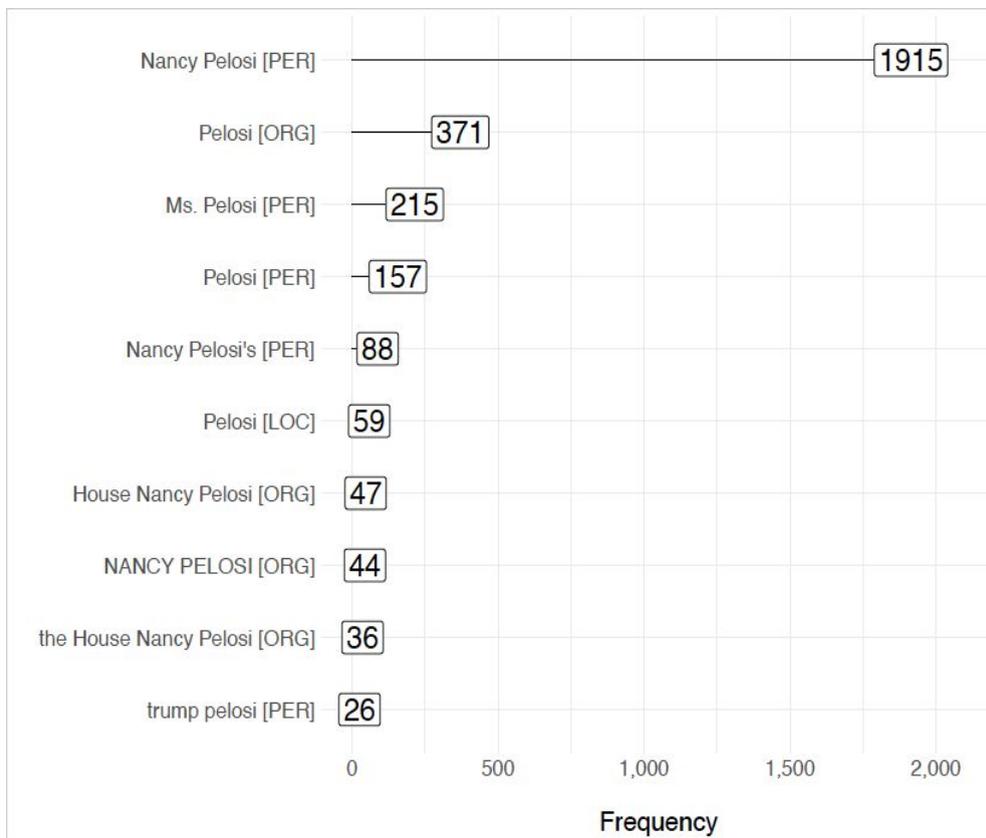

Figure 8: Frequency of Top 10 Entities Associated with "Nancy Pelosi"

Building on the above analysis, we sought to examine how often and what fraction of the time the ten most frequently-mentioned U.S. Congresspeople were correctly resolved by our ER algorithm. Figure 8 depicts the top ten ways in which "Nancy Pelosi" is resolved. "Nancy Pelosi", how-ever, is resolved a total of over 400 different ways, demonstrating room for improvement. Figure 9 depicts the number of different ways in which a Congressperson is resolved (x-axis), the cumulative sum of the number of times the name was resolved (y-axis), with a percentage, in square brackets, indicating the fraction of references attributed to the most frequent instance of the entity.

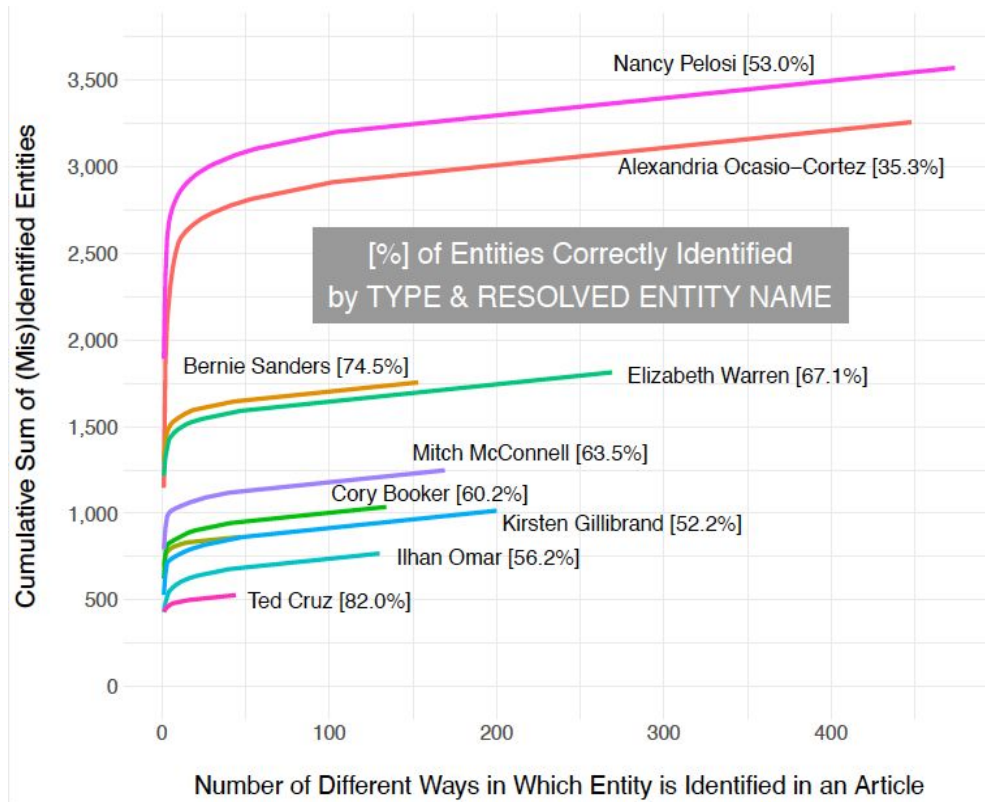

Figure 9: (Mis)Identification of Congresspeople

The many different entities for Congresspeople can be partly attributed to errors in NER models that associate extraneous characters and tokens with names. As discussed above, we also observed Congresspeople being associated with the incorrect labels of LOCATION or ORGANIZATION. Sporadic errors in the spelling of a Congressperson's name from source articles also contributed to errors in this Entity Resolution step. For example, "Alexander OcasioCortez" [sic] appears as a misspelling of the representative.

## 5. CASE STUDIES

To demonstrate the power of our resource, we choose two case studies. The first study uses the locations mentioned in an article by different news organisations to expose a location bias. The second study demonstrates how a critical event on a given day can alter the sentiment ascribed to a politician by a news organization, and how our resources provides the high level of resolution necessary to detect said changes.

### 5.1. Location Bias

In seeking to determine whether news organizations have a geographic reporting bias, we plotted all named LOCATION entities and their frequencies for the Atlantic and Slate news articles between July, 2018 and June, 2019. The geomap, produced using OpenHeatMap, is depicted in Figure 10, and demonstrates that, despite having a larger volume of articles than the Atlantic, articles found in Slate produce fewer mappable locations.

Moreover, locations referenced are concentrated on the North American coast (Eastern and Western United States), the British Islands and Southern France. Counterintuitively, the smaller volume of articles from the Atlantic produce a larger number and wider variety of references to locations.

The location bias for Slate is not wholly unexpected. The first sentence in its description on Wikipedia is, "Slate is an online magazine that covers current affairs, politics, and culture in the United States."

We find a similar pattern for the other news organizations from which we scraped data. For example, the BBC shows a plurality of articles referencing the United Kingdom, Ire-land, the USA, with several references to former British colonies (India, Australia, New Zealand, South Africa, etc.). These findings are a confirmation of [4].

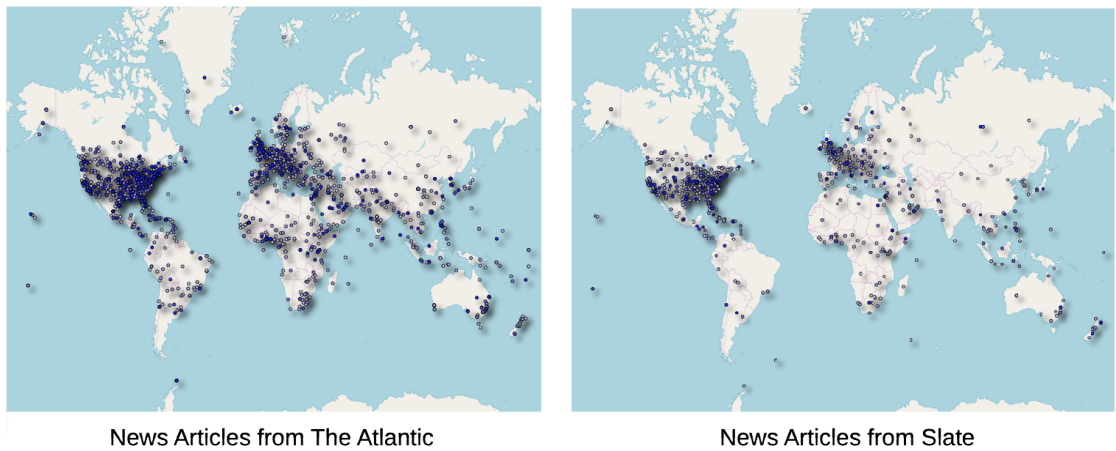

News Articles from The Atlantic       News Articles from Slate

**Figure 10: Locations Referenced by The Atlantic and Slate**

## 5.2. Sentiment for José Serrano

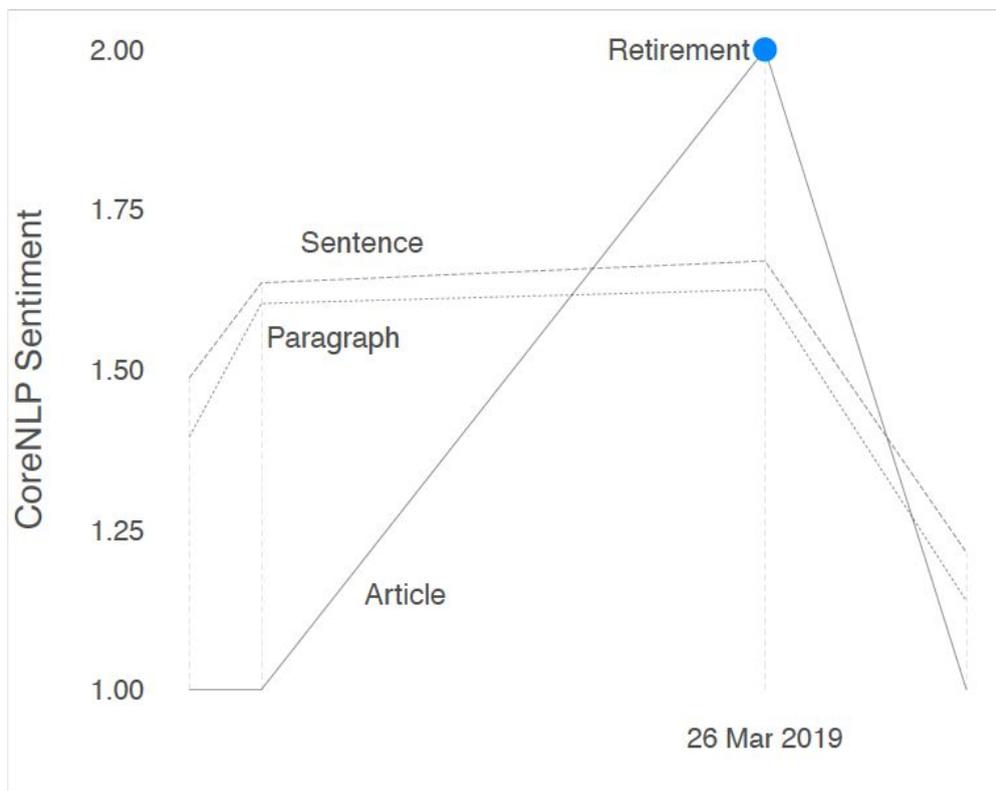

Figure 11: Sentiment for José E. Serrano at the Sentence, Paragraph and Article levels

Scouring left-leaning news organizations, we observed a peculiar pattern. When reporting on a left-leaning politicians—in America, typically, a Democrat—the sentiment associated with this reporting follows a pattern whereby the sentiment of the overall article is lower than that of the sentiment associated with paragraphs in which the politician in question is referenced, which itself has a is lower sentiment than the sentence(s) in which the politician is mentioned. In sum, the more focus there is on the politician herself, the higher the sentiment. This has been shown to be true for several left-leaning politicians when querying the Quantum Criticism corpus. This rule, however, is violated, when a seminal event. For example, when José E. Serrano, Democrat representing the 15[th] district of New York announced his retirement, the overall sentiment of the article jumped to a value higher than either the paragraph-or sentence-level sentiment (see Figure 11), a change only detectable with the sentence-level of granularity provided by the Quantum Criticism corpus.

## 3. Conclusion and Future Work

We collected a database of news articles from five popular media organizations, placed each article in a pipeline to identify named entities and determined the affect of each named entity. We identified interesting patterns and confirmed a geographic selection bias found by other re-searchers. Collecting new news data every two hours, our platform shows great promise for future research, and will further benefit from additional iterations.

We aspire to make this tool even more useful through the addition of news articles from additional news sources. Be-cause news is sometimes underreported by organizations—see

Radiolab's *Breaking Bongo* [27] for one unusual case–we will also consider adding selected tweets and other social media messages from individuals and organizations. We have already collected hundreds of thousands of candidate tweets which we have not yet filtered for relevance or made available. When coupled with better or customized tools for NER, sentiment and entity resolution, we believe this project has the potential to uncover a wide range of phenomena.

The addition of one or more frameworks for coding event data, such as CAMEO, COPDAB or others would also in-crease the usefulness of the tool. Such frameworks would allow comparison of the same set of events across different media outlets, communities and countries.

The authors acknowledge and thank culture critic Theodore Gioia, who originated the term Quantum Criticism and was a guide and inspiration for this work. Software Engineer Nikhil Barapatre led the effort to produce the web interface. The authors would also like to thank the Machine learning, Artificial and Gaming Intelligence and, Computing at Scale (MAGICS) Lab at the University of San Francisco for supporting this research with mentorship and computational infrastructure.